\documentclass[pra,superscriptaddress]{revtex4}% Physical Review B
\usepackage{graphicx}% Include figure files
\usepackage{dcolumn}% Align table columns on decimal point
\usepackage{bm}% bold math
\usepackage{amsmath}
\usepackage{amssymb,amsbsy}
\usepackage{color,ulem,cancel}
\newcommand{\beq}{\begin{equation}}
\newcommand{\eeq}{\end{equation}}
\newcommand{\bea}{\begin{eqnarray}}
\newcommand{\eea}{\end{eqnarray}}
\newcommand{\bec}{\begin{center}}
\newcommand{\enc}{\end{center}}
\newcommand{\bfr}{\begin{flushright}}
\newcommand{\efr}{\end{flushright}}
\newcommand{\la}{\langle}
\newcommand{\ra}{\rangle}

\newcommand{\om}{\omega}
\newcommand{\kap}{\kappa}
\newcommand{\gam}{\gamma}

\newcommand{\lam}{\lambda}

\newcommand{\tc}{\widetilde{c}}%\rm 
\newcommand{\tom}{\widetilde{\omega}}

\newcommand{\tkap}{\widetilde{\kappa}}
\newcommand{\tgam}{\widetilde{\gamma}}

\newcommand{\hb}{\hat{b}} 
\newcommand{\hc}{\hat{c}} 
\newcommand{\hd}{\hat{d}} 
\newcommand{\hO}{\hat{O}} 
 
\newcommand{\hH}{\hat{H}} 
 
\newcommand{\hX}{\hat{X}} 
\newcommand{\hY}{\hat{Y}}

\begin{document}
%%%%%%%%%%%%%%%%%%%%%%%%%%%%%%%%%%%%%%%%%%%%%%%%%%%%%%%%%%%%%%%%%%%%%%%%%%%%%%%
%%%%%%%%%%%%%%%%%%%%%%%%%%%%%%%%%%%%%%%%%%%%%%%%%%%%%%%%%%%%%%%%%%%%%%%%%%%%%%%
\title{
Elliptical rotation of cavity amplitude in ultrastrong waveguide QED
}
%%%%%%%%%%%%%%%%%%%%%%%%%%%%%%%%%%%%%%%%%%%%%%%%%%%%%%%%%%%%%%%%%%%%%%%%%%%%%%%
%%%%%%%%%%%%%%%%%%%%%%%%%%%%%%%%%%%%%%%%%%%%%%%%%%%%%%%%%%%%%%%%%%%%%%%%%%%%%%%
\author{Kazuki Koshino}
\affiliation{College of Liberal Arts and Sciences, Tokyo Medical and Dental
University, Ichikawa, Chiba 272-0827, Japan}
%%%%%%%%%%%%%%%%%%%%%%%%%%%%%%%%%%%%%%%%%%%%%%%%%%%%%%%%%%%%%%%%%%%%%%%%%%%%%%%
%%%%%%%%%%%%%%%%%%%%%%%%%%%%%%%%%%%%%%%%%%%%%%%%%%%%%%%%%%%%%%%%%%%%%%%%%%%%%%%
\date{\today}
%%%%%%%%%%%%%%%%%%%%%%%%%%%%%%%%%%%%%%%%%%%%%%%%%%%%%%%%%%%%%%%%%%%%%%%%%%%%%%%
%%%%%%%%%%%%%%%%%%%%%%%%%%%%%%%%%%%%%%%%%%%%%%%%%%%%%%%%%%%%%%%%%%%%%%%%%%%%%%%
\begin{abstract}
We investigate optical response of a linear waveguide 
quantum electrodynamics (QED) system, namely, 
an optical cavity coupled to a waveguide. 
Our analysis is based on exact diagonalization of the overall Hamiltonian 
and is therefore rigorous even in 
the ultrastrong coupling regime of waveguide QED. 
Owing to the counter-rotating terms in the cavity-waveguide coupling, 
the motion of cavity amplitude in the phase space is elliptical in general.
Such elliptical motion becomes remarkable in the ultrastrong coupling regime
due to the large Lamb shift comparable to the bare cavity frequency. 
We also reveal that such elliptical motion does not propagate into the output field 
and present an analytic form of the reflection coefficient
that is asymmetric with respect to the resonance frequency.
\end{abstract}
%%%%%%%%%%%%%%%%%%%%%%%%%%%%%%%%%%%%%%%%%%%%%%%%%%%%%%%%%%%%%%%%%%%%%%%%%%%%%%%
% \pacs{
% 42.50.Ar, % Photon statistics and coherence theory 
% 42.50.Pq, % Cavity quantum electrodynamics; micromasers 
% 42.65.Sf % optical spatio-temporal dynamics 
% }
\maketitle

%%%%%%%%%%%%%%%%%%%%%%%%%%%%%%%%%%%%%%%%%%%%%%%%%%%%%%%%%%%%%%%%%%%%%%%%%%%%%%%
\section{introduction}%\label{sec:intro}
%%%%%%%%%%%%%%%%%%%%%%%%%%%%%%%%%%%%%%%%%%%%%%%%%%%%%%%%%%%%%%%%%%%%%%%%%%%%%%%
Cavity quantum electrodynamics (QED) deals with 
the interaction between a single atom and 
a discretized photon mode confined in a resonator, 
which is the simplest embodiment of quantum light-matter interaction. 
The cavity QED systems have been realized in various physical platforms:
just to cite a few, single atoms coupled to an optical cavity, 
a semiconductor quantum dot in a photonic-crystal cavity, 
and a superconducting qubit coupled to a transmission-line resonator.
Interestingly, regardless of its physical platform, 
a cavity QED system is characterized by several universal parameters, 
such as $\om_a$ and $\om_c$ (atom and cavity frequencies),
$g$ (atom-photon coupling), $\kap$ (cavity decay rate), and
$\gam$ (atomic decay rate into environments). 
% and $\gam_p$ (atomic pure dephasing rate). 
%
In the history of cavity QED, extensive efforts have been made 
to reach the strong-coupling regime ($g>\kap, \gam$), 
where the vacuum Rabi oscillation and splitting become observable~\cite{sc1,sc2,sc3,sc4}. 
In usual strong-coupling systems, the coupling is still by far smaller than 
the resonance frequencies of the atom and cavity.
Recently, 
% owing largely to the emergence of circuit QED~\cite{cQED1}, 
attainments of the ultrastrong-coupling ($g \gtrsim \om_{a,c}/10$)
and deep-strong-coupling ($g \gtrsim \om_{a,c}$) regimes 
have been reported~\cite{us1,us2,us3,us4,us5,deep1,deep2}. 
In such ultrastrong-coupling systems, 
the counter-rotating terms in the Hamiltonian, 
which do not conserve the total number of excitations
and are usually negligible in the weakly coupled systems,
result in several intriguing physical phenomena, 
such as the Bloch-Siegert shift~\cite{BS1,BS2}, 
virtual photons in the ground state~\cite{vp1,vp2,vp3,vp4,vp5}, 
and the number non-conserving optical processes such as 
multiphoton vacuum Rabi oscillation~\cite{multi0,multi1,multi2}.

Waveguide QED deals with the interaction between a single atom and 
a one-dimensional continuum of photon modes, 
typically provided by a waveguide attached to the atom. 
The parameters to characterize waveguide QED systems are
$\om_a$, $\gam_e$ (atomic decay rate into waveguide)
and $\gam_i$ (atomic decay rate into environments). 
The strong-coupling regime in waveguide QED is defined by $\gam_e > \gam_i$, 
namely, the condition that radiation from the atom is dominantly forwarded 
to the waveguide~\cite{wQED1,wQED2,wQED3,wQED4,wQED5,wQED6}. 
This is reflected in spectroscopy 
as a strong suppression of transmission near the atomic resonance. 
Following the definitions in cavity QED, 
the ultra- and deep-strong waveguide QED 
should be defined as $\gam_e \gtrsim \om_a/10$ and $\gam_e \gtrsim \om_a$, respectively. 
The ultrastrong and deep-strong regimes of waveguide QED 
have already been reached using a superconducting qubit~\cite{usQED1,usQED2}. 
Theoretically, up to the usual strong-coupling regime, 
perturbative treatment of dissipation
based on the rotating-wave and Born-Markov approximations 
provides convenient and powerful theoretical tools, 
such as the Lindblad master equation and the input-output formalism~\cite{th0a,th0b}. 
However, this is not the case in highly dissipative regimes, 
and rigorous numerical methods are actively developed~\cite{th1,th2,th3,th4}. 

%------------------------------------------------------------------------------
\begin{figure}%[b]
\begin{center}
\includegraphics[width=70mm]{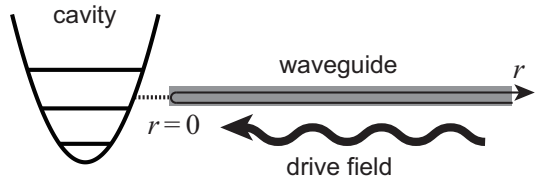}
\end{center}
\caption{
Schematic of a cavity-waveguide system. 
A cavity is coupled to a semi-infinite waveguide, 
through which a monochromatic drive field is applied. 
The $r<0$ ($r>0$) region in the waveguide
corresponds to the input (output) port. 
}
\label{fig:setup}
\end{figure}
%------------------------------------------------------------------------------

In this study, we investigate a linear waveguide QED setup, 
namely, a harmonic oscillator coupled to a waveguide, 
and investigate its optical response to a classical drive field 
applied through this waveguide. 
A merit of this system is that 
the overall Hamiltonian is diagonalizable by the Fano's method~\cite{Fano1,Fano2,Fano3}
and rigorous optical response is accessible 
even for highly dissipative situations. 
We report an elliptic motion of the oscillator in the phase space, 
which occurs, in principle, even in the usual waveguide QED setups
but becomes remarkable in the ultrastrong-coupling regime
due to the large Lamb shift. 
However, in contrast with the intuition provided by the input-output theory, 
such elliptic motion does not propagate into the waveguide. 
We also obtain an analytic formula of the reflection/transmission coefficient, 
which is asymmetric with respect to the renormalized cavity frequency.
We hope that the rigorous optical response presented here 
would be useful for developing theoretical tools applicable 
to highly dissipative cavity and waveguide QEDs.

%%%%%%%%%%%%%%%%%%%%%%%%%%%%%%%%%%%%%%%%%%%%%%%%%%%%%%%%%%%%%%%%%%%%%%%%%%%%%%%
\section{theoretical model}
\label{sec:model}
%%%%%%%%%%%%%%%%%%%%%%%%%%%%%%%%%%%%%%%%%%%%%%%%%%%%%%%%%%%%%%%%%%%%%%%%%%%%%%%
%==============================================================================
\subsection{Hamiltonian}%\label{ssec:Ham}
%==============================================================================
In a setup considered in this study (Fig.~\ref{fig:setup}), 
a cavity is coupled to a semi-infinite waveguide
and a monochromatic drive field is applied through this waveguide.
In the natural units of $\hbar=v=1$, 
where $v$ is the photon velocity in the waveguide,
the Hamiltonian of the overall system is given by
\bea
\hH &=& \om_b \hb^{\dag} \hb + \int_0^{\infty}dk 
\left[ k \hc_k^{\dag} \hc_k + \xi_k (\hb^{\dag}+\hb)(\hc_k^{\dag}+\hc_k) \right],
\label{eq:Ham}
\eea
where $\om_b$ is the {\it bare} cavity frequency, and 
$\hb$ and $\hc_k$ are the annihilation operators of the cavity mode 
and the waveguide mode with wave number $k$, respectively, 
satisfying the bosonic commutation relations, 
$[\hb, \hb^{\dag}]=1$ and $[\hc_k, \hc^{\dag}_{k'}]=\delta(k-k')$.
The cavity-waveguide coupling $\xi_k$ is a real function of $k$. 
In this study, in order that the Fano diagonalization is applicable, 
we assume the following conditions on $\xi_k$~\cite{Fano2}:
(i)~$\xi_k$ is nonzero for $k>0$, 
(ii)~$\xi^2_k$ is an odd function of $k$, namely, $\xi^2_{-k}=-\xi^2_k$,
and (iii)~the coupling is weak enough to satisfy 
\bea
\int_0^{\infty} dk \ \xi^2_k/k &<& \om_b/4.
% \verb#{eq:xi_cond}#
\label{eq:xi_cond}
\eea

%%%%%%%%%%%%%%%%%%%%%%%%%%%%%%%%%%%%%%%%%%%%%%%%%%%%%%%%%%%%%%%%%%%%%%%%%%%%%%%
\subsection{Drude-form coupling}%\label{ssec:inivec}
%%%%%%%%%%%%%%%%%%%%%%%%%%%%%%%%%%%%%%%%%%%%%%%%%%%%%%%%%%%%%%%%%%%%%%%%%%%%%%%
To be more concrete, we employ a Drude-form for the cavity-waveguide coupling, 
\bea
\xi_k^2 &=& C\frac{k}{k^2+\om_x^2}, 
\label{eq:xik2}
\eea
where $C$ is a constant % having a dimension of (frequency)$^2$, 
and $\om_x$ is the cutoff frequency. 
We assume $\om_x \gg \om_b$ so that 
the coupling is Ohmic ($\propto k$) near the cavity resonance. 
We set $\om_x=5~\om_b$ hereafter. % throughout this study.
We denote the radiative decay rate of the cavity mode into the waveguide by $\kap$. 
By naively applying the Fermi golden rule, we obtain $\kap=2\pi\xi_{\om_b}^2$.
Therefore, we set the constant $C$ as 
\bea
C &=& \frac{\kap(\om_b^2+\om_x^2)}{2\pi\om_b}. 
\label{eq:xik2C}
\eea
In this paper, we employ a dimensionless quantity $\kap/\om_b$
as a measure of the strength of the cavity-waveguide coupling.

%%%%%%%%%%%%%%%%%%%%%%%%%%%%%%%%%%%%%%%%%%%%%%%%%%%%%%%%%%%%%%%%%%%%%%%%%%%%%%%
\subsection{Renormalization of frequency and decay rate}%\label{ssec:inivec}
%%%%%%%%%%%%%%%%%%%%%%%%%%%%%%%%%%%%%%%%%%%%%%%%%%%%%%%%%%%%%%%%%%%%%%%%%%%%%%%
%------------------------------------------------------------------------------
\begin{figure}%[b]
\begin{center}
\includegraphics[width=70mm]{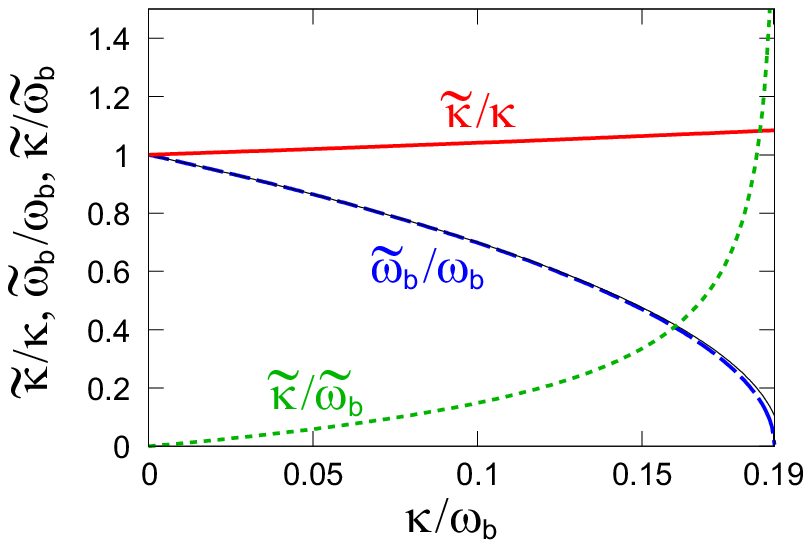}
\end{center}
\caption{
Dependences of the cavity decay rate ($\tkap/\kap$, solid) and 
resonance frequency ($\tom_b/\om_b$, dashed)
on the cavity-waveguide coupling, $\kap/\om_b$. 
Their ratio, $\tkap/\tom_b$, is plotted by a dotted line.
The ultrastrong coupling ($\tkap/\tom_b>0.1$)
is attained for $\kap/\om_b > 0.076$ and 
the deep-strong coupling ($\tkap/\tom_b>1$)
is attained for $\kap/\om_b > 0.183$. 
An alternative expression of the renormalized frequency, 
Eq.~(\ref{eq:tomb2}), is also shown (thin solid).
}
\label{fig:tomb}
\end{figure}
%------------------------------------------------------------------------------
Since the Fermi golden rule is in principle valid 
only for a weak cavity-waveguide coupling, 
$\kap$ may deviate from the actual decay rate $\tkap$,  
particularly for a stronger coupling.
Furthermore, the resonance frequency $\om_b$ also acquires a Lamb shift
and takes a renormalized value $\tom_b$. 
As we observe later in Sec.~\ref{ssec:3b}, 
$\tom_b$ and $\tkap$ are identified as
\bea
\tom_b &=& \mathrm{Re}(\lam_1),
% \verb#{eq:tombdef}#
\label{eq:tombdef}
\\
\tkap &=& 2 \ \mathrm{Im}(\lam_1),
% \verb#{eq:tkapdef}#
\label{eq:tkapdef}
\eea
where $\lam_1$ is a complex cavity frequency,   
which is a solution of the cubic equation~(\ref{eq:cubic}) 
in the first quadrant (Fig.~\ref{fig:lam123}).

From Eq.~(\ref{eq:xi_cond}), we have
% \bea
% \kap/\om_b &<& \frac{\om_b\om_x}{\om_b^2+\om_x^2}.
% \eea
$\kap/\om_b < \om_b\om_x/(\om_b^2+\om_x^2)$.
This inequality sets an upper bound for the coupling strength: 
$\kap/\om_b < 0.192$ for $\om_x=5\om_b$.
However, as we discuss in Sec.~\ref{ssec:3b}, 
from the condition that the renormalized frequency $\tom_b$ is positive, 
we have a more strict upper bound, $\kap/\om_b < 0.190$.

In Fig.~\ref{fig:tomb}, we plot the dependences of 
$\tom_b$ and $\tkap$ on $\kap/\om_b$.
We observe that, beyond the perturbative regime of $\kap/\om_b \ll 1$, 
the agreement between $\tkap$ and $\kap$ is fairly good even for stronger coupling. 
In contrast, the renormalized cavity frequency decreases drastically 
as the coupling becomes stronger. As a result, 
not only the ultrastrong coupling regime ($\tkap/\tom_b > 0.1$) 
but also the deep-strong coupling regime ($\tkap/\tom_b > 1$) is attainable
within this theoretical model.

%%%%%%%%%%%%%%%%%%%%%%%%%%%%%%%%%%%%%%%%%%%%%%%%%%%%%%%%%%%%%%%%%%%%%%%%%%%%%%%
\subsection{Initial state vector}%\label{ssec:inivec}
%%%%%%%%%%%%%%%%%%%%%%%%%%%%%%%%%%%%%%%%%%%%%%%%%%%%%%%%%%%%%%%%%%%%%%%%%%%%%%%
In this study, we investigate the optical response of a cavity
driven by a monochromatic classical field 
applied through the waveguide (Fig.~\ref{fig:setup}). 
The positively rotating part of drive amplitude is given by
\bea
E(r,t) &=& E_d e^{ik_d(r-t)},
% \verb#{eq:Ed}#
\label{eq:Ed}
\eea
where $E_d$ and $k_d$ are the complex amplitude and 
wavenumber/frequency of the drive, respectively.
At the initial moment ($t=0$), we assume that
the whole system is in the ground state expect the drive field in the waveguide, 
which is in a coherent state. 
The initial state vector is then written as
\bea
|\psi_i\ra &=& 
\exp\left(\sqrt{2\pi}E_{d}\hc_{k_{d}}^{\dag}
-\sqrt{2\pi}E^*_{d}\hc_{k_{d}}\right)|vac\ra,
\label{eq:psii}
\eea
where $|vac\ra$ is the overall ground state.

The real-space representation $\tc_r$ of the waveguide field operator 
is defined as the Fourier transform of $\hc_k$,  
\bea
\tc_r &=& \frac{1}{\sqrt{2\pi}} \int_0^{\infty} dk \ e^{ikr}\hc_k.
% \verb#{eq:tcr}#
\label{eq:tcr}
\eea
We can check that 
$\la \tc_r(0) \ra \equiv \la \psi_i|\tc_r(0)|\psi_i\ra = E(r,0)$.
Strictly speaking, 
the real-space representation of the waveguide mode
depends on the boundary condition of the waveguide at $r=0$. 
For example, for a closed boundary condition, 
the waveguide mode function takes the form of 
$f_k(r) =\sqrt{2/\pi} \sin(kr)=(ie^{-ikr}-ie^{ikr})/\sqrt{2\pi}$~\cite{norm}.
Therefore, we should add a phase factor $i$ ($-i$)  
for the input (output) port in Eq.~(\ref{eq:tcr}),  
which accounts for the sign flip upon reflection at a mirror. 
However, we employ Eq.~(\ref{eq:tcr}) 
as the real-space representation of waveguide modes
for simplicity.
This introduces no problem except for definition of the relative phase 
in the input and output ports.

%%%%%%%%%%%%%%%%%%%%%%%%%%%%%%%%%%%%%%%%%%%%%%%%%%%%%%%%%%%%%%%%%%%%%%%%%%%%%%%
\section{Diagonalization}%\label{sec:dia}
%%%%%%%%%%%%%%%%%%%%%%%%%%%%%%%%%%%%%%%%%%%%%%%%%%%%%%%%%%%%%%%%%%%%%%%%%%%%%%%
%==============================================================================
\subsection{General formula}%\label{ssec:3a}
%==============================================================================
The Hamiltonian [Eq.~(\ref{eq:Ham})] is bilinear in bosonic operators
and can be diagonalized by the Fano's method. 
When the cavity-waveguide coupling is weak enough to satisfy Eq.~(\ref{eq:xi_cond}),  
we can rewrite the Hamiltonian as
\bea
\hH &=& 
\int_0^{\infty} dk \ k \hd_k^{\dag} \hd_k, 
\label{eq:Ham2}
\eea
where $\hd_k$ is an eigenmode annihilation operator 
satisfying the bosonic commutation relation, 
\bea
[\hd_k, \hd_{k'}^{\dagger}] &=& \delta(k-k').
\label{eq:norm}
\eea
$\hd_k$ is given by linear combination of the original bosonic operators as
\bea
\hd_k &=& \beta_1(k) \hb + \beta_2(k) \hb^{\dag} + 
\int_0^{\infty} dq \left[ \gam_1(k,q) \hc_q + \gam_2(k,q) \hc_q^{\dag} \right],
\eea
where the coefficients are given by (see Appendix~\ref{app:deter} for derivation)
\bea
\beta_1(k) &=& % \frac{k+\om_b}{2\om_b \xi_k y(k)}=
\frac{(k+\om_b)\xi_k}{k^2-\om_b^2 z(k)},
\label{eq:beta1}
\\
\beta_2(k) &=& \frac{(k-\om_b)\xi_k}{k^2-\om_b^2 z(k)},
\label{eq:beta2}
\\
\gam_1(k,q) &=& \delta(k-q) + \tgam_1(k,q),
% \verb#{eq:gam1}#
\label{eq:gam1}
\\
\gam_2(k,q) &=& \frac{2\om_b \xi_k \xi_q}{(k+q)[k^2-\om_b^2 z(k)]},
% \verb#{eq:gam2}#
\label{eq:gam2}
\eea
where 
\bea
\tgam_1(k,q) &=& \frac{2\om_b \xi_k \xi_q}{(k-q-i0)[k^2-\om_b^2 z(k)]},
% \verb#{eq:tgam1}#
\label{eq:tgam1}
\eea
and $z(k)$ is a dimensionless quantity
representing the self-energy correction for the resonator frequency,
\bea
z(k) &=& 1 + \frac{2}{\om_b}\int_{-\infty}^{\infty}dq \frac{\xi_q^2}{k-q-i0}.
% \verb#{eq:z(k)}#
\label{eq:z(k)}
\eea
Inversely, the bare operators $\hb$ and $\hc_k$ are expressed 
in terms of the eigenoperators by
\bea
\hb &=& \int_0^{\infty} dq [\beta_1^*(q) \hd_q - \beta_2(q) \hd_q^{\dag}],
% \verb#{eq:hb}#
\label{eq:hb}
\\
\hc_k &=& \int_0^{\infty} dq [\gam_1^*(q,k)\hd_q - \gam_2(q,k)\hd_q^{\dag}].
% \verb#{eq:hck}#
\label{eq:hck}
\eea

%==============================================================================
\subsection{Specific results for Drude-form coupling}
\label{ssec:3b}
%==============================================================================
%------------------------------------------------------------------------------
\begin{figure}
\begin{center}
\includegraphics[width=70mm]{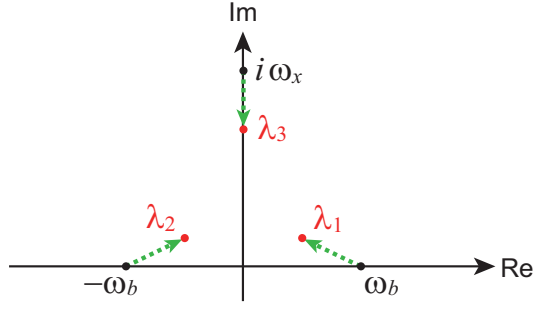}
\end{center}
\caption{
$\lam_{1,2,3}$ on the complex plane. Arrows indicate the directions
as the cavity-waveguide coupling $\kap$ is increased.
}
\label{fig:lam123}
\end{figure}
%------------------------------------------------------------------------------
When the cavity-waveguide coupling takes the Drude form
[Eqs.~(\ref{eq:xik2}) and (\ref{eq:xik2C})], 
$z(k)$ and $k^2-\om_b^2 z(k)$ are rewritten as follows,
\bea
z(k) &=& 1+\frac{2\pi i C}{\om_b(k-i\om_x)}, 
\\
k^2-\om_b^2 z(k) &=& \frac{(k-\lam_1)(k-\lam_2)(k-\lam_3)}{k-i\om_x}, 
\label{eq:k2omb2}
\eea
where $\lam_{1,2,3}$ are the solutions of the following cubic equation for $k$, 
\bea
k^3-i\om_x k^2 -\om_b^2 k + (i\om_x\om_b^2 - 2i\pi C \om_b) &=& 0.
% \verb#{eq:cubic}#
\label{eq:cubic}
\eea
As shown in Fig.~\ref{fig:lam123}, 
$\lam_1$ ($\lam_2$) is on the first (seond) quadrant % $\lam_2=-\lam_1^*$, 
and $\lam_3$ is on the positive imaginary axis.
The real and imaginary parts of $\lam_1$ correspond to
the Lamb-shifted resonance frequency $\tom_b$ and
half of the decay rate $\tkap/2$ [Eqs.~(\ref{eq:tombdef}) and (\ref{eq:tkapdef})].
For reference, 
we present the perturbative solution of Eq.~(\ref{eq:cubic})
with respect to the cavity-waveguide coupling $\kap$. 
The zeroth-order solutions are 
$\lam_1^{(0)}=\om_b$, $\lam_2^{(0)}=-\om_b$, and $\lam_3^{(0)}=i\om_x$. 
Up to the first order in $\kap$, the three solutions are given by 
% \bea
% \lam_1 & \approx & 
% \left(\om_b -\frac{\kap\om_x}{2\om_b}\right) + i\frac{\kap}{2},
% \\
% \lam_2 & \approx & 
% -\left(\om_b -\frac{\kap\om_x}{2\om_b}\right) + i\frac{\kap}{2},
% \\
% \lam_3 & \approx & 
% i\om_x - i\kap.
% \eea
$\lam_1 \approx (\om_b-\kap\om_x/2\om_b) + i\kap/2$, 
$\lam_2 \approx -(\om_b-\kap\om_x/2\om_b) + i\kap/2$, 
and $\lam_3 \approx i\om_x - i\kap$.

For an extremely strong coupling, 
$\lam_1$ and $\lam_2$ also become purely imaginary. 
% The condition that $\lam_1$ ($\lam_2$) remains in the first (second) quadrant, 
The condition that 
the renormalized frequency $\tom_b$ remain positive, 
in other words, 
$\lam_1$ and $\lam_2$ are not purely imaginary, 
is that $\kap < [\om_b^2\om_x-f(\mu_-)]/(\om_b^2+\om_x^2)$, 
where $f(x)=x^3-\om_x x^2+\om_b^2 x$ and 
$\mu_-$ is a smaller root of the $df/dx=0$, 
namely, $\mu_- = (\om_x-\sqrt{\om_x^2-3\om_b^2})/3$. 
For $\om_x = 5~\om_b$, this condition is $\kap/\om_b < 0.190$.

%%%%%%%%%%%%%%%%%%%%%%%%%%%%%%%%%%%%%%%%%%%%%%%%%%%%%%%%%%%%%%%%%%%%%%%%%%%%%%%
\section{optical response}%\label{sec:op}
%%%%%%%%%%%%%%%%%%%%%%%%%%%%%%%%%%%%%%%%%%%%%%%%%%%%%%%%%%%%%%%%%%%%%%%%%%%%%%%
%===========================================================================
\subsection{Cavity Amplitude}
\label{ssec:camp}
\begin{figure} % [b]
\begin{center}
\includegraphics[width=150mm]{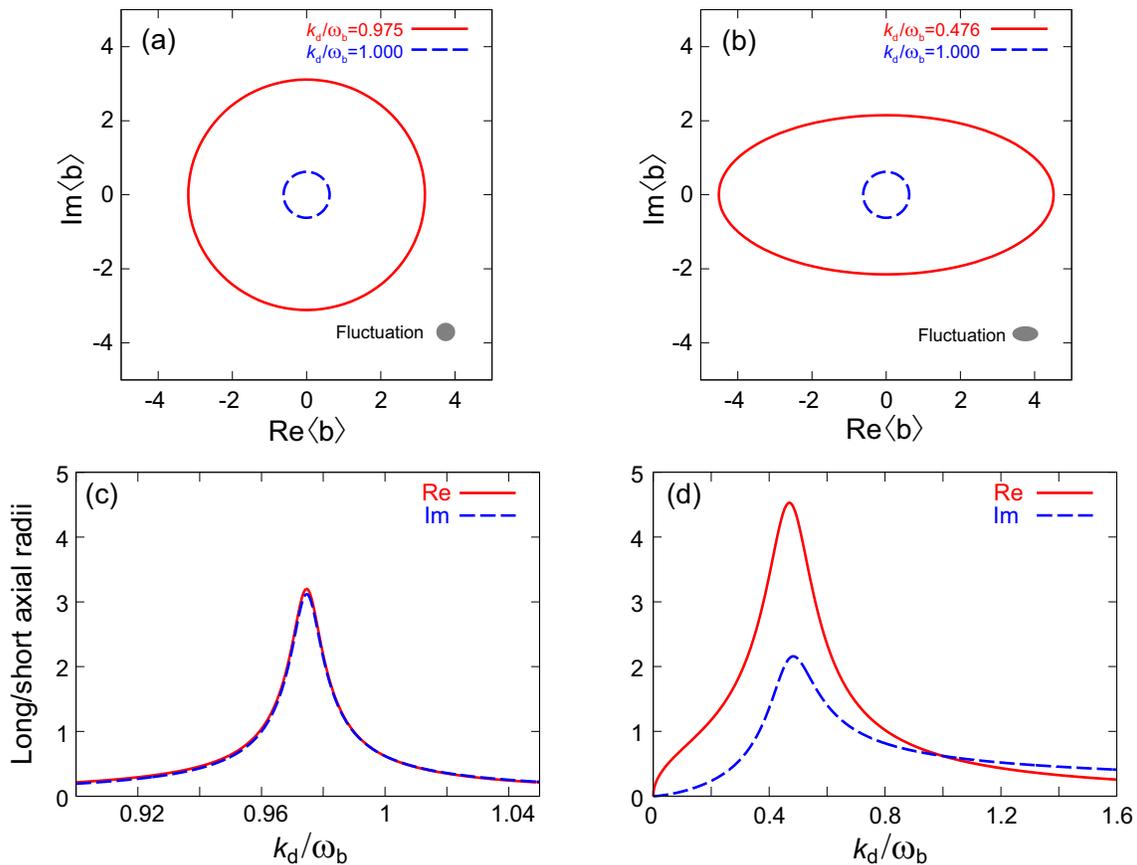}
\end{center}
\caption{
Elliptical motion of the cavity amplitude.
(a)~Trajectories on the phase space for $\kap/\om_b=0.01$.
The drive frequency is set at 
the renormalized resonance $\tom_b(=0.975~\om_b)$ (solid) and
the bare resonance $\om_b$ (dashed).
The photon rate of the drive field is set at $|E_d|^2=2.5~\kap$, 
at which the mean intra-cavity photon number is estimated to be 
$\la \hb^{\dag}\hb \ra = 4|E_d|^2/\kap =10$ on resonance, 
following the input-output theory.
The uncertainty ellipse is also shown.
(b)~The same plot as (a) for $\kap/\om_b=0.15$. 
The renormalized resonance is $\tom_b=0.476~\om_b$. 
(c)~Dependence of the long (solid line) and short (dotted line) axial radii
on the drive frequency $k_d$ for $\kap/\om_b=0.01$. 
(d)~The same plot as (c) for $\kap/\om_b=0.15$. 
}
\label{fig:tra}
\end{figure}

%===========================================================================
In this section, we investigate 
time evolution of the whole system 
from the initial state vector, Eq.~(\ref{eq:psii}).  
We first observe the amplitude of the cavity mode, 
$\la \hb(t) \ra \equiv \la \psi_i|\hb(t)|\psi_i\ra$.
Since $\hd_k$ is an eigenoperator of the Hamiltonian, 
$\hb(t)$ is given, from Eq.~(\ref{eq:hb}), by
\bea
\hb(t) &=& \int_0^{\infty}dq \left[
e^{-iqt} \beta_1^*(q) \hd_q - e^{iqt} \beta_2(q) \hd^{\dag}_q
\right].
\label{eq:hbt}
\eea
Furthermore, $|\psi_i\ra$ is an eigenstate of $\hd_q$ and satisfies
\bea
\hd_q |\psi_i\ra &=&
\sqrt{2\pi} [E_d \gam_1(q,k_d) + E^*_d \gam_2(q,k_d)] |\psi_i\ra.
% \verb#{eq:dqpsi}#
\label{eq:dqpsi}
\eea
From these results, $\la \hb(t) \ra$ is given by
\bea
\la \hb(t) \ra 
&=&
\sqrt{2\pi}E_d \int_0^{\infty} dq \left[ 
e^{-iqt}\beta_1^*(q) \gam_1(q,k_d ) - e^{iqt}\beta_2(q) \gam_2^*(q,k_d )
\right]
\nonumber
\\
&+&
\sqrt{2\pi}E^*_d \int_0^{\infty} dq \left[ 
e^{-iqt}\beta_1^*(q) \gam_2(q,k_d ) - e^{iqt}\beta_2(q) \gam_1^*(q,k_d )
\right].
\eea
This is divided into stationary and transient components as
$\la \hb(t) \ra=\la \hb(t) \ra_s + \la \hb(t) \ra_t$. 
The stationary component is given by
\bea
\la \hb(t) \ra_s &=&
\sqrt{2\pi} \beta_1^*(k_d) E_d  e^{-ik_d t} -
\sqrt{2\pi} \beta_2(k_d) E_d ^* e^{ik_d t}. 
% \verb#{eq:bts}#
\label{eq:bts}
\eea
The transient component is presented in Appendix~\ref{app:tra}.
Putting $E_d =|E_d |e^{i\theta_d}$, we have
\bea
\mathrm{Re}\la \hb(t) \ra_s &=&
\sqrt{8\pi}|E_d|\om_b\xi_{k_d}
\mathrm{Re}
\left(
\frac{e^{i(k_d t-\theta_d)}}{k_d^2-\om_b^2 z(k_d)}
\right),
\label{eq:Re}
\\
\mathrm{Im}\la \hb(t) \ra_s &=&
-\sqrt{8\pi}|E_d|k_d \xi_{k_d}
\mathrm{Im}
\left(
\frac{e^{i(k_d t-\theta_d)}}{k_d^2-\om_b^2 z(k_d)}
\right).
\label{eq:Im}
\eea
These equations indicate that
the motion of the cavity amplitude $\la \hb(t) \ra_s$ 
on the phase space is elliptical in general; 
the ratio of the vertical (imaginary) radius 
relative to the horizontal (real) radius 
is $k_d/\om_b$, and thus depends on the drive frequency.
However, such elliptical motion is not remarkable 
when the cavity-waveguide coupling $\kap$ is small. 
For a small $\kappa$ case, 
strong optical response is obtained within a narrow frequency region 
around the renormalized cavity frequency $\tom_b$,  
which is close to the bare frequency $\om_b$. 
For example, when $\kap/\om_b=0.01$, 
the renormalized frequency amounts to $\tom_b = 0.975~\om_b$ [Eq.~(\ref{eq:tomb2})]. 
Therefore, the motion is almost circular for small $\kap$, 
as we observe in Figs.~\ref{fig:tra}~(a) and (c). 
In contrast, for a large $\kappa$ case, 
the motion on the phase space becomes highly elliptical, 
as we observe in Figs.~\ref{fig:tra}~(b) and (d). 
This is due to the large frequency renormalization (Lamb shift). 
When $\kap/\om_b=0.15$, the renormalized frequency amounts to $\tom_b = 0.476~\om_b$. 
%---------------------------------------------------------------------------
%---------------------------------------------------------------------------

%===========================================================================
\subsection{Quadrature Fluctuations}%\label{sec:sH}
%===========================================================================
%---------------------------------------------------------------------------
\begin{figure} % [b]
\begin{center}
\includegraphics[width=70mm]{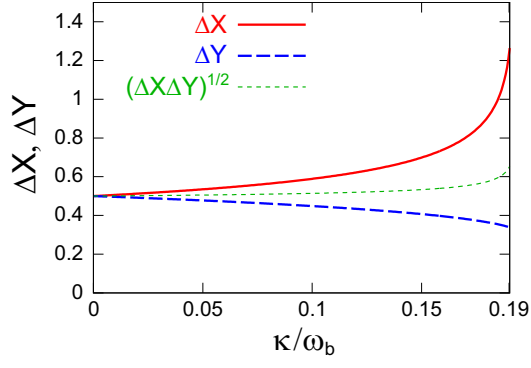}
\end{center}
\caption{Quadrature fluctuations:
$\Delta X$ (solid), $\Delta Y$ (dashed), 
and $\sqrt{\Delta X \Delta Y}$ (thin dotted).
$\om_x/\om_b=5$. 
}
\label{fig:DxDy}
\end{figure}
%---------------------------------------------------------------------------
Here, we investigate the quadrature fluctuations of the cavity mode. 
We define the $\hX$ and $\hY$ quadratures 
by $\hX=(\hb+\hb^{\dagger})/2$ and $\hY=-i(\hb-\hb^{\dagger})/2$, respectively, 
and their fluctuations by
% $\la (\Delta \hX)^2 \ra= \la \hX^2 \ra - \la \hX \ra^2$ and 
% $\la (\Delta \hY)^2 \ra= \la \hY^2 \ra - \la \hY \ra^2$, respectively, 
$\Delta X=\sqrt{\la \hX^2 \ra - \la \hX \ra^2}$ and 
$\Delta Y=\sqrt{\la \hY^2 \ra - \la \hY \ra^2}$, respectively, 
where $\la \hO \ra = \la \psi_i|\hO|\psi_i\ra$.
From these definitions, we have 
\bea
\Delta X
&=& 
\frac{\sqrt{1 + 2\la \hb^{\dag}(t), \hb(t)\ra + 2\mathrm{Re}\la \hb(t), \hb(t) \ra}}{2},
\label{eq:DX2}
\\
\Delta Y
&=& 
\frac{\sqrt{1 + 2\la \hb^{\dag}(t), \hb(t)\ra - 2\mathrm{Re}\la \hb(t), \hb(t) \ra}}{2},
\label{eq:DY2}
\eea
% \bea
% \la (\Delta \hX)^2 \ra 
% &=& 
% [1 + 2\la \hb^{\dag}(t), \hb(t)\ra + 2\mathrm{Re}\la \hb(t), \hb(t) \ra]/4,
% \label{eq:DX2}
% \\
% \la (\Delta \hY)^2 \ra 
% &=& 
% [1 + 2\la \hb^{\dag}(t), \hb(t)\ra - 2\mathrm{Re}\la \hb(t), \hb(t) \ra]/4,
% \label{eq:DY2}
% \eea
where $\la \hO, \hO' \ra \equiv \la \hO\hO' \ra-\la \hO \ra \la \hO' \ra$. 
From Eqs.~(\ref{eq:hbt}) and (\ref{eq:dqpsi}), 
we can confirm that
both $\la \hb^{\dag}(t), \hb(t) \ra$ and $\la \hb(t), \hb(t) \ra$
% $\la \hb^{\dag}(t), \hb(t) \ra = \la \hb^{\dag}(t) \hb(t) \ra - |\la \hb(t) \ra|^2 $ and 
% $\la \hb(t), \hb(t) \ra = \la \hb^2(t) \ra - \la \hb(t) \ra^2$
reduces to the following time-independent quantities,  
\bea
\la \hb^{\dag}, \hb \ra 
&=&
% \la \hb^{\dag}(t) \hb(t) \ra - |\la \hb(t) \ra|^2 = 
\int_0^{\infty}dq |\beta_2(q)|^2,
% \verb#{eq:<b*,b>}#
\label{eq:<b*,b>}
\\
\la \hb, \hb \ra &=&
% \la \hb^2(t) \ra - \la \hb(t) \ra^2 = 
-\int_0^{\infty}dq \beta_1^*(q) \beta_2(q),
% \verb#{eq:<b,b>}#
\label{eq:<b,b>}
\eea
and that the quadrature fluctuations, $\Delta X$ and $\Delta Y$,
are identical to those of the vacuum fluctuations.
The integrals appearing in Eqs.~(\ref{eq:<b*,b>}) and (\ref{eq:<b,b>}) can be performed 
analytically for the Drude-form coupling (Appendix~\ref{app:C}).

Figure~\ref{fig:DxDy} plots the dependences of $\Delta X$ and $\Delta Y$
on the cavity-waveguide coupling $\kap$. 
We observe that there exists squeezing in $Y$ quadrature, 
and the degree of squeezing increases for larger $\kap$. 
The state is not a minimum uncertainty state, 
since $\sqrt{\Delta X \Delta Y}>1/2$ as we observe in Fig.~\ref{fig:DxDy}.

%===========================================================================
\subsection{Amplitude of waveguide field}%\label{sec:sH}
%===========================================================================
% Here, we investigate the amplitude of the waveguide field. 
From Eqs.~(\ref{eq:hck}) and (\ref{eq:dqpsi}), 
the amplitude of the waveguide field in the wavenumber representation is given by
\bea
\la \hc_k(t) \ra &=&
\sqrt{2\pi}E_d \int_0^{\infty} dq \left[ 
e^{-iqt}\gam_1^*(q,k) \gam_1(q,k_d ) - e^{iqt}\gam_2(q,k) \gam_2^*(q,k_d )
\right]
\nonumber
\\
&+&
\sqrt{2\pi}E^*_d \int_0^{\infty} dq \left[ 
e^{-iqt}\gam_1^*(q,k) \gam_2(q,k_d ) - e^{iqt}\gam_2(q,k) \gam_1^*(q,k_d )
\right]. 
\eea
Using Eqs.~(\ref{eq:gam1})--(\ref{eq:tgam1}), this quantity is rewritten as follows,
\bea
\la \hc_k(t) \ra &=&
\sqrt{2\pi}E_d  \left[
e^{-ik_d t}\delta(k-k_d ) + e^{-ikt}\tgam_1(k,k_d ) + e^{-ik_d t}\tgam_1^*(k_d ,k)
\right]
\nonumber
\\
&-& i\sqrt{2/\pi} \om_b \xi_k \xi_{k_d } E_d  
\int_{-\infty}^{\infty}dq \frac{e^{-iqt}}{(q-k+i0)(q-k_d -i0)}
\left(
\frac{1}{q^2-\om_b^2 z(q)}-\frac{1}{q^2-\om_b^2 z^*(q)}
\right)
\nonumber
\\
&+& \sqrt{2\pi}E_d ^* \left[
e^{-ikt}\gam_2(k,k_d ) - e^{ik_d t}\gam_2(k_d ,k)
\right]
\nonumber
\\
&+& i\sqrt{2/\pi} \om_b \xi_k \xi_{k_d } E_d ^* 
\int_{-\infty}^{\infty}dq \frac{e^{iqt}}{(q+k-i0)(q-k_d +i0)}
\left(
\frac{1}{q^2-\om_b^2 z(q)}-\frac{1}{q^2-\om_b^2 z^*(q)}
\right).
% \verb#{eq:<c_k(t)>}#
\label{eq:<c_k(t)>}
\eea
The integral in the second line in the above equation
can be performed by employing the residue theorem. 
The integrand has four poles in the lower complex plane of $q$
at $k-i0$, $\lam_1^*$, $\lam_2^*$, and $\lam_3^*$, 
and the latter three poles yield transient components.
Therefore, the stationary component of the second line 
comes from the pole at $k-i0$ and is given by 
$-\sqrt{8\pi}\om_b\xi_k\xi_{k_d } \frac{E_d e^{-ikt}}{k-k_d -i0}
(\frac{1}{k^2-\om_b^2 z(k)}-\frac{1}{k^2-\om_b^2 z^*(k)})$.
% This partly cancels with $\sqrt{2\pi}E_d  e^{-ikt}\tgam_1(k,k_d )$. 
Repeating the same arguments, 
the stationary component of the fourth line of Eq.~(\ref{eq:<c_k(t)>}) is given by
$-\sqrt{8\pi}\om_b\xi_k\xi_{k_d } \frac{E_d ^*e^{-ikt}}{k+k_d }
(\frac{1}{k^2-\om_b^2 z(k)}-\frac{1}{k^2-\om_b^2 z^*(k)})$.
% This partly cancels with $\sqrt{2\pi}E_d ^* e^{-ikt}\gam_2(k,k_d )$. 
As a result, the stationary component of % $\la \hc_k(t) \ra$ 
the waveguide amplitude is written as
\bea
\la c_k(t) \ra &=& 
\la c_k(t) \ra^{(1)} + \la c_k(t) \ra^{(2)} + \la c_k(t) \ra^{(3)},
% \verb#{eq:<c_k(t)>v3}#
\label{eq:<c_k(t)>v3}
\\
\la c_k(t) \ra^{(1)} &=&
\sqrt{2\pi}\delta(k-k_d ) E_d e^{-ik_d t},  
\\
\la c_k(t) \ra^{(2)} &=&
\frac{\sqrt{8\pi}\om_b \xi_k \xi_{k_d } E_d }{k-k_d -i0}
\left(
\frac{e^{-ikt}}{k^2-\om_b^2 z^*(k)} - 
\frac{e^{-ik_d t}}{k_d ^2-\om_b^2 z^*(k_d )}
\right),
% \verb#{eq:<ckt>2}#
\label{eq:<ckt>2}
\\
\la c_k(t) \ra^{(3)} &=& 
\frac{\sqrt{8\pi}\om_b \xi_k \xi_{k_d } E_d ^*}{k+k_d }
\left(
\frac{e^{-ikt}}{k^2-\om_b^2 z^*(k)} - 
\frac{e^{ik_d t}}{k_d ^2-\om_b^2 z(k_d )}
\right).
\eea

%---------------------------------------------------------------------------
\begin{figure}
\begin{center}
\includegraphics[width=130mm]{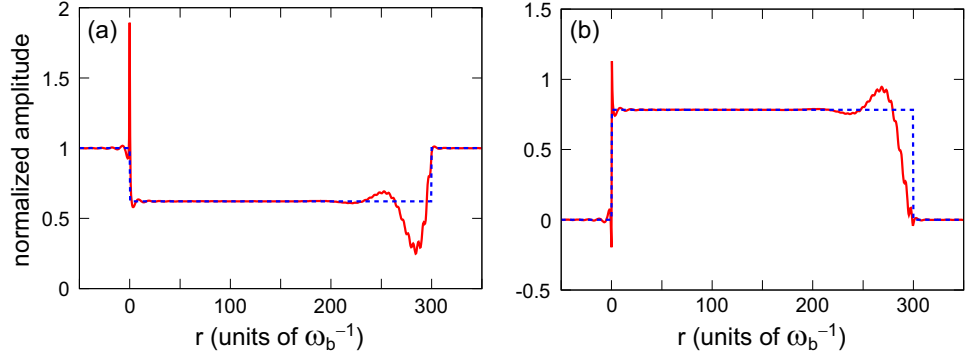}
\end{center}
\caption{
Normalized amplitude of the waveguide field, $\la \tc_r(t) \ra/E(r,t)$.
(a)~Real and (b)~imaginary parts.
Solid lines represent the rigorous numerical results, 
and dotted lines represent the approximate one given by Eq.~(\ref{eq:<tcr>}).
The parameters are chosen as follows:
$\kap/\om_b=0.1$, $k_d/\om_b=0.6$, and $t=300/\om_b$.
}
\label{fig:crt2}
\end{figure}
%---------------------------------------------------------------------------

We switch to the real-space representation, $\la \tc_r(t) \ra$,  
using Eq.~(\ref{eq:tcr}). 
$\la \tc_r(t) \ra^{(1)}$ is immediately given by
\bea
\la \tc_r(t) \ra^{(1)} &=& E_d  e^{ik_d (r-t)}.
\eea
Obviously, this is nothing but the input drive field of Eq.~(\ref{eq:Ed}). 
Regarding $\la \tc_r(t) \ra^{(2)}$, 
the principal contribution comes from the pole at $k=k_d+i0$
in the right-hand-side of Eq.~(\ref{eq:<ckt>2}).
Therefore, we can employ the following approximation, 
$\la c_k(t) \ra_s^{(2)} \approx 
\sqrt{8\pi}\om_b \xi_{k_d }^2 E_d 
[k_d ^2-\om_b^2 z^*(k_d )]^{-1} [k-k_d -i0]^{-1}
\left( e^{-ikt} - e^{-ik_d t} \right)$.
Then, we have
\bea
\la \tc_r(t) \ra^{(2)} & \approx & 
-\frac{4\pi i \om_b \xi_{k_d }^2}{k_d ^2-\om_b^2 z^*(k_d )}
\theta(r)\theta(t-r) E_d  e^{ik_d (r-t)},
\eea
where $\theta$ is the Heaviside step function.
This represents the radiation from the cavity 
emitted into the positive $r$ region. 
Finally, $\la \tc_r(t) \ra^{(3)}$ yields no propagating wave. 
Combining these results, 
we obtain the following analytic form of $\la \tc_r(t) \ra$:
\bea
\la \tc_r(t) \ra & \approx & 
\left(
1-\frac{4\pi i \om_b \xi_{k_d }^2}{k_d ^2-\om_b^2 z^*(k_d )}
\theta(r)\theta(t-r)
\right) \times E_d  e^{ik_d (r-t)}.
% \verb#{eq:<tcr>}#
\label{eq:<tcr>}
\eea
% This equation is similar to the one derivable from the standard input-output theory, 
% except that the coefficient of the cavity radiation takes a more complicated form. 

The spatial shape of $\la \tc_r(t) \ra$ is plotted in Fig.~\ref{fig:crt2}, 
in which the rigorous shape [numerical Fourier transform of Eq.~(\ref{eq:<c_k(t)>v3})] 
is plotted by solid lines
and the approximate form [Eq.~(\ref{eq:<tcr>})] 
is plotted by dotted lines. 
We observe good agreement between them, 
except the deviations at the wavefront of the cavity radiation ($r \lesssim t$)
and at the cavity position ($r \sim 0$). 
The former deviation originates in the transient cavity response, 
which is not taken into account in Eq.~(\ref{eq:<tcr>}). 
The transient response vanishes within a timescale of $\kap^{-1}$, 
which agrees with our observation in Fig.~\ref{fig:crt2}.
On the other hand, 
the latter deviation around the cavity position originates in the fact that 
the cavity-waveguide interaction has a finite bandwidth in the wavenumber space
and therefore is not spatially local in the present theoretical model. 
The bandwidth of the cavity-waveguide coupling 
is of the order of $\om_b$ in the wavenumber space, 
and is therefore of the order of $\om_b^{-1}$ in the real space. 
This explains the deviation localized at the origin in Fig.~\ref{fig:crt2}. 

A notable fact is that, 
in contrast with the intracavity field amplitude [Eq.~(\ref{eq:bts})]
that is composed of both positively and negatively oscillating components,
the waveguide field amplitude in the output port [Eq.~(\ref{eq:<tcr>})]
is composed only of the positively oscillating one.
Therefore, the elliptic motion is specific to the intracavity amplitude.
%
% This comes from the fact that 
% the cavity amplitude oscillating with the negative 
% does not propagate into the output port.

%===========================================================================
\subsection{Refection coefficient}%\label{ssec:ref}
\begin{figure}
\begin{center}
\includegraphics[width=70mm]{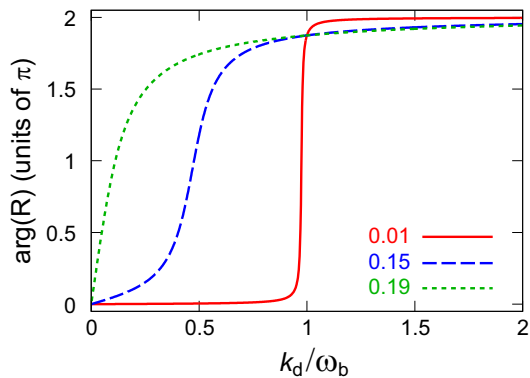}
\end{center}
\caption{
Phase shift upon reflection as a function of the drive frequency.
The cavity-waveguide coupling strength, $\kap/\om_b$, is indicated. 
}
\label{fig:ref}
\end{figure}

%===========================================================================
The refection coefficient is identified as
$R = \la \tc_r(t) \ra/E(r,t)$ at the output port ($r>0$). 
From Eq.~(\ref{eq:<tcr>}), $R$ is identified as
\bea
R(k_d ) &=& 1-\frac{4\pi i \om_b \xi_{k_d }^2}{k_d ^2-\om_b^2 z^*(k_d )}.
% \verb#{eq:Ref}#
\label{eq:Ref}
\eea
We can check that $|R|=1$ for any input frequency $k_d $. 
This implies that input field is reflected completely coherently, 
which is characteristic to linear optical response. 
In Fig.~\ref{fig:ref}, we plot 
the phase shift upon reflection, $\arg R$, 
as a function of the drive frequency $k_d$, 
varying the cavity-waveguide coupling. 
As we increase the coupling, 
we observe the broadening of the linewidth
and the redshift of the resonance frequency. 
The spectrum takes a kink-shaped form around the renormalized frequency. 
For a weak coupling, the spectrum is anti-symmetric 
with respect to the renormalized frequency, 
as is predicted by standard input-output theory.
However, for a stronger coupling, such symmetry is gradually lost.

We can determine the renormalized resonance frequency $\tom_b$
as the drive frequency achieving the $\pi$ phase shift, $R(\tom_b)=-1$. 
From this condition, $\tom_b$ is analytically given by
\bea
\tom_b^2 &=& 
\frac{\om_b^2-\om_x^2+\sqrt{(\om_b^2 + \om_x^2)(\om_b^2 + \om_x^2 -4\kap\om_x)}}{2}.
% \verb#{eq:tomb2}#
\label{eq:tomb2}
\eea
As we can confirm in Fig.~\ref{fig:tomb}, 
this is almost identical to the former 
definition of $\tom_b$ by Eq.~(\ref{eq:tombdef}). 
We observe in Fig.~\ref{fig:ref} that 
the reflection coefficient becomes independent of the coupling strength
$\kap/\om_b$ at the bare cavity resonance, $k_d =\om_b$; 
we can check that $R(\om_b)=(\om_x-i\om_b)/(\om_x+i\om_b)$. 

%---------------------------------------------------------------------------
%---------------------------------------------------------------------------

%===========================================================================
\subsection{Open waveguide}%\label{ssec:ref}
\begin{figure}
\begin{center}
\includegraphics[width=140mm]{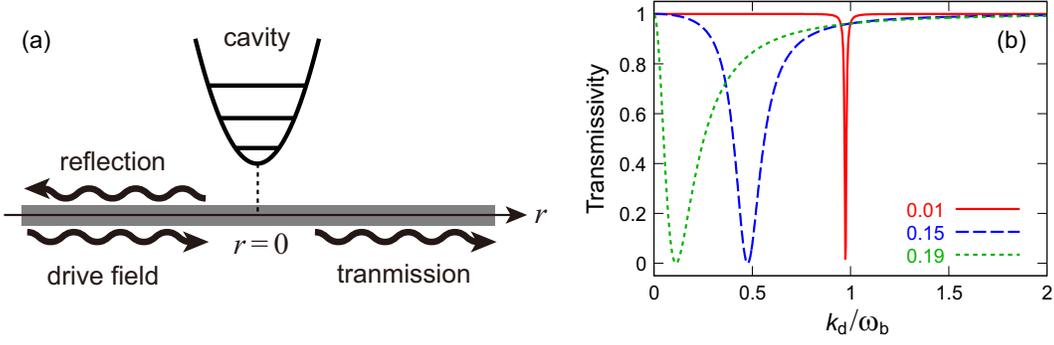}
\end{center}
\caption{
(a)~Schematic of a cavity coupled to an open waveguide. 
(b)~Transmissiveity $|T'|^2$ as a function of the drive frequency.
The cavity-waveguide coupling strength, $\kap/\om_b$, is indicated. 
}
\label{fig:open}
\end{figure}

%===========================================================================
In the previous subsection, 
we have determined the reflection coefficient $R$
when a semi-infinite waveguide is coupled to a cavity (Fig.~\ref{fig:setup}). 
From this result, we can readily determine
the reflection and transmission coefficients $R'$ and $T'$,
when the cavity is coupled to an open waveguide [Fig.~\ref{fig:open}(a)]. 
The amplitude of waveguide field in this case is written as 
\bea
E(r,t) &=& E_d e^{-i\om_d t} \times 
\begin{cases}
e^{ik_dr} + R' e^{-ik_dr} & (r<0)
\\
T' e^{ik_dr} & (0<r) 
\end{cases}.
\eea
We divide this field into even and odd components. 
The even component interacts with the cavity
whereas the odd component does not.
The even component is defined by $E_s(r,t)=[E(r,t)+E(-r,t)]/2$
and is therefore given by 
$E_s(r,t)=\frac{1}{2}E_d e^{-ik_d(r+t)} + \frac{R'+T'}{2}E_d e^{ik_d(r-t)}$ for $r>0$. 
Since the first (second) term in the right-hand-side of this equation
represents the incoming (outgoing) field, 
we have $R'+T'=R$.
Similarly, 
the odd component is defined by $E_a(r,t)=[E(r,t)-E(-r,t)]/2$
and is therefore given by 
$E_s(r,t)=-\frac{1}{2}E_d e^{-ik_d(r+t)} + \frac{T'-R'}{2}E_d e^{ik_d(r-t)}$ for $r>0$. 
Since the incoming field simply transmits the cavity without interaction, 
we have $T'-R'=1$. Therefore,
\bea
R' &=& (R-1)/2,
\\
T' &=& (R+1)/2.
\eea
We can readily confirm that $|R'|^2+|T'|^2=1$. 
The transmissivity $|T'|^2$ is plotted in Fig.~\ref{fig:open}(b)
as a function of the drive frequency.
We observe that the symmetric transmission dip for a weak coupling case (solid line) 
gradually becomes asymmetric 
as the cavity-waveguide coupling increases (dashed and dotted lines).

%---------------------------------------------------------------------------
%---------------------------------------------------------------------------

%%%%%%%%%%%%%%%%%%%%%%%%%%%%%%%%%%%%%%%%%%%%%%%%%%%%%%%%%%%%%%%%%%%%%%%%%%%%%%%
\section{summary} %\label{sec:summary}
%%%%%%%%%%%%%%%%%%%%%%%%%%%%%%%%%%%%%%%%%%%%%%%%%%%%%%%%%%%%%%%%%%%%%%%%%%%%%%%
In this study, we investigated optical response of 
a linear waveguide QED system, namely, 
an optical cavity coupled to a waveguide. 
Our analysis is based on exact diagonalization of the overall Hamiltonian, 
and is therefore rigorous % regardless of the strength of the cavity-waveguide coupling. 
even in the ultrastrong and deep-strong coupling regimes of waveguide QED, 
in which the perturbative treatments of dissipation
such as the Lindblad master equation are no longer valid. 
We observed that the motion of the cavity amplitude 
in the phase space is elliptical in general,
owing to the counter-rotating terms in the cavity-waveguide coupling. 
Such elliptical motion becomes remarkable 
in the ultrastrong coupling regime 
due to the large Lamb shift of the cavity frequency 
comparable to its bare frequency. 
However, such an elliptical motion of the cavity amplitude 
is not reflected in the output field, 
contrary to the intuition by the input-output theory. 
We obtained an analytic expression of the reflection/transmission coefficient, 
which becomes asymmetric with respect to the resonance frequency
as the cavity-waveguide coupling is increased.

%%%%%%%%%%%%%%%%%%%%%%%%%%%%%%%%%%%%%%%%%%%%%%%%%%%%%%%%%%%%%%%%%%%%%%%%%%%%%%%
\section*{Acknowledgments}
%%%%%%%%%%%%%%%%%%%%%%%%%%%%%%%%%%%%%%%%%%%%%%%%%%%%%%%%%%%%%%%%%%%%%%%%%%%%%%%
The author acknowledges fruitful discussions with T. Shitara and I. Iakoupov. 
This work is supported in part by 
JST CREST (Grant No. JPMJCR1775), JST ERATO (Grant no. JPMJER1601), 
MEXT Q-LEAP, and JSPS KAKENHI (Grant No. 19K03684). 

\appendix
%%%%%%%%%%%%%%%%%%%%%%%%%%%%%%%%%%%%%%%%%%%%%%%%%%%%%%%%%%%%%%%%%%%%%%%%%%%%%%%
% \section{determination of $\beta_{1,2}$ and $\gam_{1,2}$}
\section{Fano diagonalization}
\label{app:deter}
%%%%%%%%%%%%%%%%%%%%%%%%%%%%%%%%%%%%%%%%%%%%%%%%%%%%%%%%%%%%%%%%%%%%%%%%%%%%%%%
From Eqs.~(\ref{eq:Ham2}) and (\ref{eq:norm}), we have $[\hd_k, \hH] = k\hd_k$.
This leads the following equations:
\bea
(k-\om_b)\beta_1(k) &=& \int_0^{\infty} dq \xi_q [\gam_1(k,q)-\gam_2(k,q)],
\label{eq:Ap1}
\\
(k+\om_b)\beta_2(k) &=& \int_0^{\infty} dq \xi_q [\gam_1(k,q)-\gam_2(k,q)],
\label{eq:Ap2}
\\
(k-q)\gam_1(k,q) &=& \xi_q [\beta_1(k)-\beta_2(k)],
\label{eq:Ap3}
\\
(k+q)\gam_2(k,q) &=& \xi_q [\beta_1(k)-\beta_2(k)].
\label{eq:Ap4}
\eea
From Eqs.~(\ref{eq:Ap2}) and (\ref{eq:Ap4}), we obtain 
$\beta_2(k)=\frac{k-\om_b}{k+\om_b}\beta_1(k)$ and 
$\gam_2(k,q)=\frac{k-q}{k+q}\gam_1(k,q)$.
Then, Eqs.~(\ref{eq:Ap1}) and (\ref{eq:Ap3}) are rewritten as
\bea
(k-\om_b)\beta_1(k) &=& 2 \int_0^{\infty} dq  \frac{q\xi_q}{k+q} \gam_1(k,q),
\label{eq:Ap5}
\\
(k-q)\gam_1(k,q) &=&  \frac{2\om_b}{k+\om_b}\beta_1(k)\xi_q.
\label{eq:Ap6}
\eea
Equation~(\ref{eq:Ap6}) is rewritten as 
\bea
\gam_1(k,q) &=&  \frac{2\om_b}{k+\om_b}\beta_1(k)\xi_q 
\left(
\frac{1}{k-q-i0}+y(k)\delta(k-q)
\right),
\eea
where $y(k)$ is a quantity to be determined. 
Substituting the above equation into Eq.~(\ref{eq:Ap5}), and using 
$\int_0^{\infty} \frac{q\xi_q^2}{(k+q)(k-q-i0)}
=\frac{1}{2}\int_{-\infty}^{\infty} \frac{\xi_q^2}{k-q-i0}$, $y(k)$ is given by
\bea
y(k) &=& \frac{1}{\xi_k^2}\left(
\frac{k^2-\om_b^2}{2\om_b}-\Sigma(k)
\right), 
\\
\Sigma(k) &=& \int_{-\infty}^{\infty}dq \frac{\xi_q^2}{k-q-i0}.
\eea
Note that $\Sigma(k)$ is the self-energy of the cavity,
satisfying $\Sigma(-k)=\Sigma^*(k)$ and $\mathrm{Im}\Sigma(k)=\pi\xi_k^2$.

Up to here, we derived the expressions of $\beta_2$, $\gam_1$ and $\gam_2$ 
in terms of $\beta_1$. 
$\beta_1(k)$ is determined by the normalization condition, Eq.~(\ref{eq:norm}).
This is rewritten as
$\delta(k-k') = \beta_1(k)\beta_1^*(k')-\beta_2(k)\beta_2^*(k')
+ \int_0^{\infty} dq [\gam_1(k,q)\gam_1^*(k',q)-\gam_2(k,q)\gam_2^*(k',q)]$,
which leads to $\frac{2\om_b \xi_k}{(k+\om_b)}|\beta_1(k)||y(k)|=1$.
By adequately choosing the phase of $\beta_1$, we obtain Eq.~(\ref{eq:beta1}),
\bea
\beta_1(k) &=& \frac{k+\om_b}{2\om_b \xi_k y(k)}=
\frac{(k+\om_b)\xi_k}{k^2-\om_b^2 z(k)}.
\eea
$\beta_2$, $\gam_1$ and $\gam_2$ are obtained accordingly.

%%%%%%%%%%%%%%%%%%%%%%%%%%%%%%%%%%%%%%%%%%%%%%%%%%%%%%%%%%%%%%%%%%%%%%%%%%%%%%%
\section{transient component of cavity mode}
\label{app:tra}
%%%%%%%%%%%%%%%%%%%%%%%%%%%%%%%%%%%%%%%%%%%%%%%%%%%%%%%%%%%%%%%%%%%%%%%%%%%%%%%
Here we present the transient component of 
the cavity amplitude, $\la\hb(t)\ra_t$, 
which is omitted in Sec.~\ref{ssec:camp}: 
\bea
\la \hb(t) \ra_t &=&
\sqrt{8\pi}E_d \om_b\xi_{k_d } \int_{-\infty}^{\infty}dq 
\frac{e^{-iqt}(q+\om_b)\xi_q^2}{(q-k_d -i0)[q^2-\om_b^2 z(q)][q^2-\om_b^2 z^*(q)]}
\nonumber
\\
&-& \sqrt{8\pi}E_d ^* \om_b\xi_{k_d } \int_{-\infty}^{\infty}dq 
\frac{e^{iqt}(q-\om_b)\xi_q^2}{(q-k_d +i0)[q^2-\om_b^2 z(q)][q^2-\om_b^2 z^*(q)]}. 
\eea
Using $\frac{\om_b\xi_q^2}{[q^2-\om_b^2 z(q)][q^2-\om_b^2 z^*(q)]}=\frac{1}{4i\pi}
(\frac{1}{q^2-\om_b^2 z(q)}-\frac{1}{q^2-\om_b^2 z^*(q)})$
and that $\frac{1}{q^2-\om_b^2 z(q)}$ has no poles on the lower half plane, 
transient component is rewritten as
\bea
\la b(t) \ra_t 
&=& 
\frac{i E_d  \xi_{k_d }}{\sqrt{2\pi}}
\int_{-\infty}^{\infty}dq \frac{e^{-iqt}(q+\om_b)}{(q-k_d -i0)[q^2-\om_b^2z^*(q)]}
+
\frac{i E_d ^* \xi_{k_d }}{\sqrt{2\pi}}
\int_{-\infty}^{\infty}dq \frac{e^{iqt}(q-\om_b)}{(q-k_d +i0)[q^2-\om_b^2z(q)]}.
\eea

%%%%%%%%%%%%%%%%%%%%%%%%%%%%%%%%%%%%%%%%%%%%%%%%%%%%%%%%%%%%%%%%%%%%%%%%%%%%%%%
\section{Integrals in Eqs.~(\ref{eq:<b*,b>}) and (\ref{eq:<b,b>})}
\label{app:C}
%%%%%%%%%%%%%%%%%%%%%%%%%%%%%%%%%%%%%%%%%%%%%%%%%%%%%%%%%%%%%%%%%%%%%%%%%%%%%%%
Here, we derive an analytical form of the integral  
in the right-hand-side of Eq.~(\ref{eq:<b*,b>}).
% which we denote by $I$. 
From Eq.~(\ref{eq:z(k)}), we have $z(k)-z^*(k)=4i\pi\xi_k^2/\om_b$.
Therefore, the integral is rewritten as
\bea
\int_0^{\infty} dq |\beta_2(q)|^2 
&=& \frac{1}{4i\pi\om_b}\left(
\int_0^{\infty} dq \frac{(q-\om_b)^2}{q^2-\om_b^2 z(q)}-\mathrm{c.c.}
\right).
% \verb#{eq:I}#
\label{eq:I}
\eea
We denote the integrand in the right-hand-side of Eq.~(\ref{eq:I}) by $f(q)$. 
Using Eq.~(\ref{eq:k2omb2}), $f(q)$ is rewritten as
\bea
f(q) &=& \frac{(q-\om_b)^2}{q^2-\om_b^2 z(q)}
= \frac{(q-\om_b)^2(q-i\om_x)}{(q-\lam_1)(q-\lam_2)(q-\lam_3)}
= 1 + \sum_{j=1}^3 \frac{c_j}{q-\lam_j}, 
% \verb#{eq:f(q)}#
\label{eq:f(q)}
\eea
where $c_j$ is a residue of $f(q)$ at $q=\lam_j$. 
Substituting Eq.~(\ref{eq:f(q)}) into Eq.~(\ref{eq:I}), 
we obtain
\bea
\int_0^{\infty} dq |\beta_2(q)|^2 
&=& 
-\frac{1}{2\pi\om_b}\sum_{j=1}^3 \mathrm{Im}\{c_j \log(-\lam_j)\}.
\eea
Repeating the same argument, 
the integral appearing in Eq.~(\ref{eq:<b,b>}) is given by
\bea
-\int_0^{\infty} dq \beta_1^*(q) \beta_2(q) 
&=& 
% \frac{1}{4i\pi\om_b}\left(
% \int_0^{\infty} dq \frac{\om_b^2-q^2}{q^2-\om_b^2 z(q)}-\mathrm{c.c.}
% \right) =
-\frac{1}{2\pi\om_b}\sum_{j=1}^3 \mathrm{Im}\{
d_j \log(-\lam_j)\},
\eea 
where $d_j$ is a residue at $q=\lam_j$ of the following function $g(q)$, 
\bea
g(q) 
&=& 
% \frac{(\om_b^2-q^2)}{q^2-\om_b^2 z(q)} = 
\frac{(\om_b^2-q^2)(q-i\om_x)}{(q-\lam_1)(q-\lam_2)(q-\lam_3)}.
% = 1 + \sum_{j=1}^3 \frac{c_j}{q-\lam_j}
% \verb#{eq:g(q)}#
\label{eq:g(q)}
\eea

%%%%%%%%%%%%%%%%%%%%%%%%%%%%%%%%%%%%%%%%%%%%%%%%%%%%%%%%%%%%%%%%%%%%%%%%%%%%%

%%%%%%%%%%%%%%%%%%%%%%%%%%%%%%%%%%%%%%%%%%%%%%%%%%%%%%%%%%%%%%%%%%%%%%%%%%%%%

%%%%%%%%%%%%%%%%%%%%%%%%%%%%%%%%%%%%%%%%%%%%%%%%%%%%%%%%%%%%%%%%%%%%%%%%%%%%%%
%%%%%%%%%%%%%%%%%%%%%%%%%%%%%%%%%%%%%%%%%%%%%%%%%%%%%%%%%%%%%%%%%%%%%%%%%%%%%%

\begin{thebibliography}{99}
\bibitem{norm}
The mode function is normalized so as to satisfy 
$\int_0^{\infty} dr f_k(r) f_{k'}(r)=\delta(k-k')$.
\end{thebibliography}
\end{document}